\documentclass[runningheads]{llncs}
\usepackage{graphicx}

\usepackage[ruled,vlined]{algorithm2e}
\usepackage{algorithmic}

\usepackage{color,soul}

\begin{document}
\title{Comparing graph data science libraries for querying and analysing datasets: \\towards data science queries on graphs\thanks{This work was  funded by the Quasimodo action in U. Orléans and the DOING action of the GDR MADICS.}}
\titlerunning{Towards data science queries on graph}
%
\author{Genoveva Vargas-Solar\inst{1} \and
Pierre Marrec\inst{2} \and
Mirian Halfeld Ferrari Alves\inst{3}}
\authorrunning{G. Vargas-Solar et al.}
%
\institute{French Council of Scientific Research, LIRIS, France \\
\email{genoveva.vargas-solar@liris.cnrs.fr} \and
Ecole Natoinal Supérieur de Lyon\\
\email{pierre.marrec@ens-lyon.fr} \and
University of Orléans, LIFO, France\\
\email{mirian@univ-orleans.fr}}
\maketitle              
\begin{abstract}

This paper presents an experimental study to
to compare analysis tools with management systems for querying and analysing graphs. Our experiment compares classic graph navigational operations queries where analytics tools and management systems adopt different execution strategies. Then, our experiment addresses data science pipelines with clustering and prediction models applied to graphs. In this kind of experiment, we underline the interest of combining both approaches and the interest of relying on a parallel execution platform for executing queries.

\end{abstract}
%
%
%
\section{Introduction}

Vast collections of heterogeneous data containing observations of phenomena have become the backbone of scientific, analytic and forecasting processes for addressing problems in domains like Connected Enterprise, Digital Mesh, and Internet-connected things and Knowledge networks. Observations can be structured as networks that have interconnection rules determined by the variables (i.e., attributes) characterising each observation. 

The graph concept is a powerful mathematical concept with associated operations that can be implemented through efficient data structures and exploited by applying different algorithms. Note that relations among observations and interconnection rules are often not explicit, and it is the role of the analytics process to deduce, discover and eventually predict them. Take, for example, a graph-based representation of the plot of the famous Saga Game of Thrones.  In this graph, it is possible to (i) Ask simple queries like the number of characters of the Saga?; (ii)
build communities to know which are the build communities the houses in which some characters are organised and how influential they are?; (iii)	compute the popularity of characters and observe its evolution; (iv)
	build maps to describe the geographical distribution of the countries; (v) be more ambitious and predict who can be the next final King or Queen?

If we group the querying techniques, we can do it across two families.  The first one concerns querying as we know it in comics and information research. Here the principle is that pipelines explore and analyse the data to profile it quantitatively and with the objective of either modelling, prediction or recommendation. In the first case, the results have a notion of completeness and probabilistic approximation. While in the other family, the results have an associated degree of error, and they may be data and queries or data samples. In this paper, we tackled exploratory queries that tackle data collections that are expanding or where the structure provides little knowledge about the data. These queries run step-by-step like pipelines, and the tasks often apply statistical, probabilistic, or data mining and artificial intelligence processing functions. Methodologies are still to come to integrate data management with the execution of algorithms that are often greedy. Of course, we are not the only ones interested in this type of query in https://www.overleaf.com/project/6136432e28f08785fdc175bcits design and execution.

Existing technology, including graph stores with different models and properties and querying facilities and analytics libraries with built-in graph analytics algorithms, provide tools for exploring graphs. The question is, which conditions the different solutions are better adapted to address different analytics queries. This paper describes an experimental approach for profiling and identifying existing tools' characteristics and how they are different and complementary. 

This paper presents an experimental study to
to compare the graph analysis approaches with the graph management and query approach. We show that the purely analytical approach achieves better execution's performance than the data management system approach for relatively small datasets. Besides, we created a model that predicts the future interactions of the characters from The Song of Ice and Fire with learning tools.

Networkx allows graph processing rather than graph database management. The graphs have a dictionary architecture, in particular, to store strings and put attributes on the edges. Oriented and non-oriented graphs are two different objects in this package. We used notebook environments to code in Python with Networkx. Initially, the notebook was hosted in Kaggle.
This comparison aims to see the difference in performance on data analysis functions between a package used to analyse graphical databases (Networkx) and software that allows the management of these databases (Neo4j).

The remainder of the paper is organized as follows.
Section \ref{sec:relatedwork} discusses related work. 
Section  \ref{sec:approach} describes our study strategy including the experimental settings, datasets and discusses the obtained results. Section \ref{sec:conclusion} concludes the paper and discusses  future work.


\section{Related Work} \label{sec:relatedwork}
\begin{figure*}[h]
   \centering
   \includegraphics[width=\linewidth]{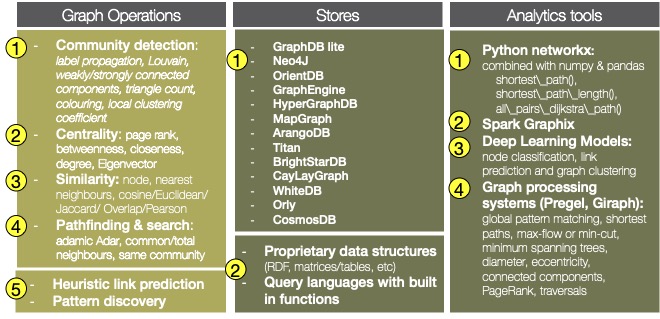}
    \caption{Graph management and processing systems}
   \label{fig:graphsystems}
 \end{figure*}

The most classic solutions are graph stores and systems that provide built-in graph operations organised in two families as shown in Figure \ref{fig:graphsystems}. Those systems implement well-known graph operations like community detection (shown in number one in Figure \ref{fig:graphsystems}) centrality where we find, for example, page rank and betweenness (number 2 in the Figure \ref{fig:graphsystems}), similarity (number 3 in the Figure \ref{fig:graphsystems}) and pathfinding and search like standard networks.

 The second family is based on data mining and machine learning techniques (shown in number 5 in Figure \ref{fig:graphsystems}) like heuristic link prediction and pattern discovery. Stores provide persistence support. There are many prominent commercial systems shown in number 1 in Figure \ref{fig:graphsystems}. They provide proprietary data structures and more or less declarative query languages with built-in functions like the ones presented before. Finally, analytics tools provide also solutions for processing graphs. For example, Python Networkx showed in number 1 in Figure \ref{fig:graphsystems}, Spark Graphix (number 2 in Figure \ref{fig:graphsystems}), deep learning models for node classification, link prediction and graph clustering (number 3 in Figure \ref{fig:graphsystems}) and graph processing systems like Pregel and Giraph.
\section {Experimentally comparing the execution of data science operations on graphs on DBMS or all-in-one programs}
\label{sec:approach}

Our work aims to study the difference in performance on data analysis functions between a package that is used to analyse graphical databases (Networkx) and software that allows the management of these databases (Neo4j) when they are used for defining data science pipelines.

\subsection{Graph and experimental setting}
For the dataset, we have chosen the data of the five books of the saga {\em The song of ice and fire} which has been extensively studied and represented in graph form.
It is an epic novel, and as such, the characters are organised in houses represented by kings and queens who are lords or ladies of the regions.
Knights engage in battles in different places, and of course, there are deaths in these battles.

\begin{figure*}[h]
   \centering
   \includegraphics[width=\linewidth]{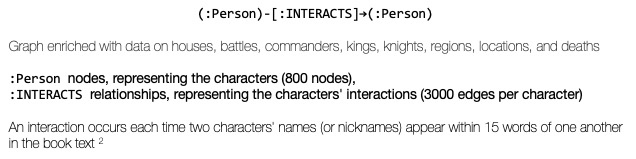}
    \caption{Game of Thrones Graph}
   \label{fig:got-graph}
 \end{figure*}

Characters of The Wise The song of Ice and Fire created by Andrew Beveridge \cite{beveridge2016network}. To create this dataset, he looked in the books to see which characters appear within 15 words of each other to determine the degree of interaction. By adapting it to our needs, we got a graph with about 800 nodes and 3000 relations. This adaptation allows for a graph that is not too big to keep the calculation times decent (see Figure \ref{fig:got-graph}). 

We selected five functions that are implemented on Neo4j and Networkx: 
\begin{itemize}
    \item {\em Page rank} measures the importance of each node within a graph, based on the number of incoming relationships and the importance of the corresponding source nodes. The underlying assumption is that a page is only as important as the pages that link to it. We assumed Neo4j and Networkx used the same implementation of these algorithms as in \cite{brin1998anatomy}. We configured the damping= 0.85 and max iterations=20 in both environments.
    \item {\em Betweenness centrality} is a way of detecting the amount of influence of a node over the flow of information in a graph. It is often used to find nodes that serve as a bridge from a subgraph to another. The algorithm calculates unweighted shortest paths between all pairs of nodes in a graph. Each node receives a score based on the number of shortest paths that pass through the node. Nodes that more frequently lie on shortest paths between other nodes will have higher betweenness centrality scores.

    \item {\em Label propagation}  finds communities in a graph using the graph structure alone as its guide and does not require a pre-defined objective function or prior information about the communities.  The intuition behind the algorithm is that a single label can quickly become dominant in a densely connected group of nodes but will have trouble crossing a sparsely connected region. Labels will get trapped inside a densely connected group of nodes, and those nodes that end up with the same label when the algorithms finish can be considered part of the same community. Both environments implement the same version of the algorithm\footnote{\url{https://neo4j.com/docs/graph-data-science/current/algorithms/label-propagation/}}.
    
    \item {\em Breadth-First Search} is a graph traversal algorithm that, given a start node, visits nodes in order of increasing distance.  Multiple termination conditions are supported for the traversal, based on either reaching one of several target nodes, reaching a maximum depth, exhausting a given budget of traversed relationship cost, or just traversing the whole graph. The output of the procedure contains information about which nodes were visited and in what order.
    
    \item {\em Minimum Spanning Tree} is a kind of pathfinding algorithm. It starts from a given node and finds all its reachable nodes and the relationships that connect them with the minimum possible weight. Prim’s algorithm \cite{choi2012prim} is one of the simplest and best-known minimum spanning tree algorithms. 
\end{itemize}
The study focused on the execution times by minimizing the delays that do not depend on the algorithm.

\begin{figure*}[h]
   \centering
   \includegraphics[width=\linewidth]{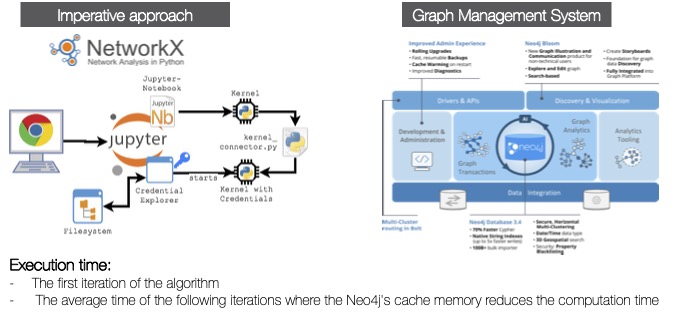}
    \caption{Graph Analytics Execution Environments}
   \label{fig:architectures}
 \end{figure*}

Since we were looking at the way the environments execute the pipelines that analyse the graphs, it is important to compare both approaches from an architectural point of view that determines execution conditions (see Figure \ref{fig:architectures}).
In the case of Python, when we use Jupyter notebooks. A client machine with a browser has access to a file system that holds the data, and it has access to a Jupyter server that has access to Python interpreters. At runtime, the data is all loaded into RAM; this poses some resource management problems. 

 For the Networkx Python environment we built the following graph. The specification of the graphh in Neo4J is given within the expression of the query.
 \begin{verbatim}
     import networkx as nx
     G=nx.Graph(name="Game of Networks")
     n=len(table['Source'])
     for i in range(n):
         G.add_edge(table['Source'][i],table['Target'][i],
                    weight=table['weight'][i])
 \end{verbatim}

The following expressions compare the code used in Python programs and Cypher to express data science queries, exploring the graph and answering these queries.

\begin{itemize}
    \item {\em Q-1: Which are the most influencial characters of the novel? }
    \begin{itemize}
        \item Cypher expression:
    \begin{verbatim}
         CALL gds.alpha.betweenness.stream({
               nodeQuery: 'MATCH (p) RETURN id(p) AS id',
               relationshipQuery: 'MATCH (p1)-[]-(p2) 
               RETURN id(p1) AS source, id(p2) AS target'})
         YIELD nodeId,centrality
         return gds.util.asNode(nodeId).name 
                as user, centrality
                order by centrality DESC limit 1
    \end{verbatim}
    
    \item Python program using Networkx method {\sf\small nx.betweenness\_centrality(G)}.
    \begin{verbatim}
        list=[]
        for i in range(100):
            a=time()
            nx.betweenness\_centrality(G)
            b=time()
            liste.append(b-a)
    \end{verbatim}
     \end{itemize}
\end{itemize}
  Of course, Python promotes imperative query programming, assuming that the underlying infrastructure provides enough main memory space to retrieve the graph and process it.  In the case of Neo4J, the preparation of main memory allocation, the tuning of specific parameters of the algorithm like the number of iterations, the precision objective to define a termination condition must be executed before the code shown above. Neo4J also works with graphs in main memory when applying data science functions. The graphs are views of persistent graphs defined using Cypher. The view can provide a subset of nodes respecting some restriction, and relations can be directed/labelled or not. It is up to the programmer to store the view and results upon the termination of the process.
  
 \paragraph{Results}
In the case of Neo4J, the graphs are stored and can be queried declaratively, but when applying graph analysis functions, the system requires the user to manage the memory and the routing of the graph pieces to the execution space.
So we compared their behaviour concerning execution time: particularly the time cost of the first iteration of the algorithms, and then calculating the average execution time for the following iterations to look at the advantage of having a cache in the case of Neo4J no-cache in the case of Python.

\begin{figure*}[h]
   \centering
   \includegraphics[width=\linewidth]{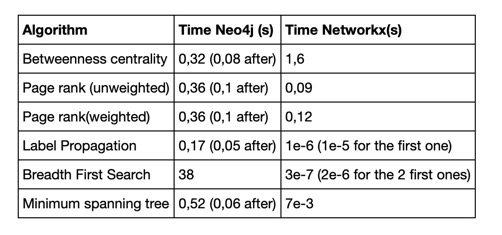}
    \caption{Comparison of execution results}
   \label{fig:exp-pipeline}
 \end{figure*}

\subsection{Graph Data Science Pipeline}
 We implemented data science queries as pipelines that combine graph matching queries and aggregations (see Figure \ref{fig:exp-pipeline}). The first group of tasks includes resource estimation (main memory) and data preparation. The second group of tasks include
exploratory, modelling and prediction operations and results assessment. However,  data preparation has not been considered in our performance comparison since we are interested in comparing data science operations execution.
Using this pipeline, we could solve a set of analytics questions by implementing notebooks in Python and Cypher queries.

\begin{figure*}[h]
   \centering
   \includegraphics[width=\linewidth]{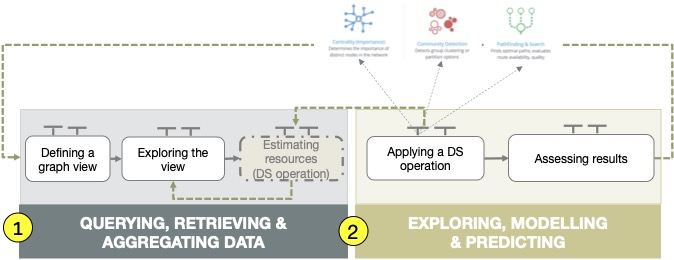}
    \caption{Graph Analytics General Pipeline}
   \label{fig:exp-pipeline}
 \end{figure*}

\begin{itemize}
 \item {Q$_1$ - {\em Which are the houses that challenge the thrones and how influential are they?}}
This question was answered applying centrality algorithms namely  page rank and betweeness centrality. 
 \item {Q$_2$ - {\em Which are the most popular characters?}}
This question was answered applying centrality algorithms namely  page rank and betweeness centrality.

 \item {Q$_3$ - {\em Which are the houses that challenge the thrones?
}}
We used the community detection family to answer this question.

 \item {Q$_4$ - {\em Who are the leading characters in Game of Thrones?
}}
The notoriety of characters was analysed with the breadth first search and minimum spanning tree.
\end{itemize}

For Networkx, we performed many tests, so the uncertainty is low. 
There are significant order-of-magnitude differences for specific algorithms, such as the width path. Indeed, if we did not limit the maximum depth to 5, the algorithm did not finish (or its execution was very long).
Afterwards, it is expected that there is a difference because Neo4j also manages the graph in real-time, whereas for Networkx, we had to recreate the graph each time we launched the notebook. 
\subsection{Link prediction}
\begin{figure*}[h]
   \centering
   \includegraphics[width=\linewidth]{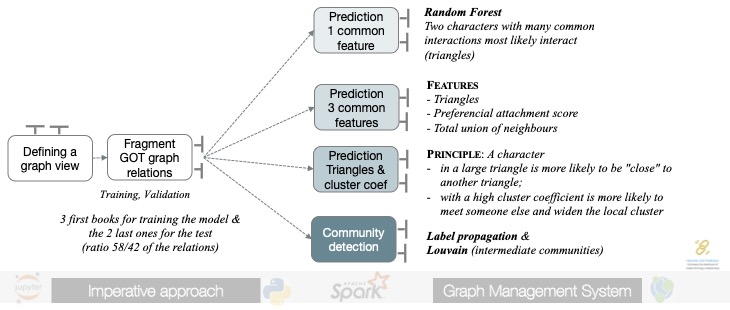}
    \caption{Link prediction pipeline}
   \label{fig:link-pipeline}
 \end{figure*}

For the link prediction part integration of two platforms with a parallel programming model with Spark \footnote{\url{https://github.com/gevargas/doing-graph-datascience-queries}}.
So, for the prediction, we compared different strategies by including properties of characters represented by the node to discover links that would be hidden.
Secret relations to beat a king or conquer a house.
So our pipeline developed different complementary branches with richer analysis to discover as many new relations as possible. Figure \ref{fig:link-pipeline} shows the general pipeline implemented for discovering links among the novel characters according to different sets of attributes.

 \paragraph{Data Preparation}
 The dataset is divided into 2 parts. The first contained about 60\% of the data and served as a learning set. The second contains 40\% of the data, and it is used to test the model's performance.

Each of the two sets comprises a certain number of pairs of nodes connected by an edge and the same number of nodes not connected by an edge. Provided that, in general, there are far fewer existing edges than possible edges, the number of unconnected node pairs in each set had to be reduced beforehand.

 \paragraph{Specifying characteristics}
We created characteristics for the nodes in the graph to correctly classify the edges so that our forest of trees could correctly.
The choice of these characteristics is the tricky part of link prediction because the suitable characteristics depend a lot on the graph's topology. These characteristics are often values calculated by a graph analysis algorithm such as Page Rank but can also be more specific functions such as the number of neighbours in common. A good feature is a value that allows the model to classify pairs of vertices correctly.

\begin{figure*}[h]
   \centering
   \includegraphics[width=\linewidth]{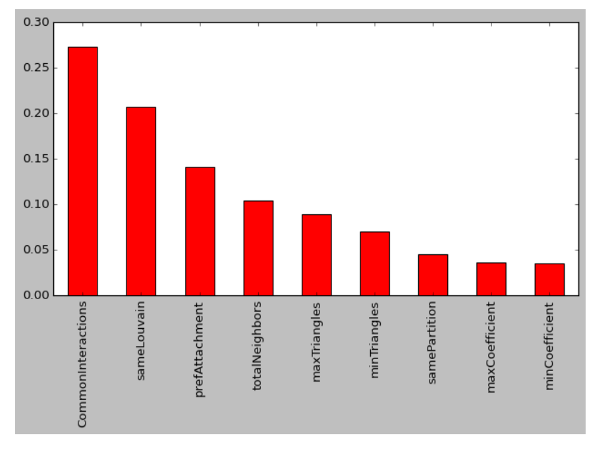}
    \caption{First model characteristics}
   \label{fig:model}
 \end{figure*}

 \paragraph{Model 1: predicting links using the attribute number of neighbours}
 We started by predicting links with the number of neighbours in common as the only characteristic as a criterion. As seen in Figure \ref{fig:model}  characters with few neighbours in common are pretty unlikely to interact with each other, and characters with many neighbours in common interact with each other. This observation gives a good clue about the usefulness of this characteristic in differentiating between interactions that will and will not exist.
After training the model with just this characteristic, 
the model is already much better than a random classifier.

 \paragraph{Model 2: predicting links according to several characters characteristics}
We  added 8 characteristics looking for a model with better prediction performance:
\begin{enumerate}
    \item The number of neighbours in common.
    \item The number of neighbours.
    \item The preferential link score which is a coefficient calculated by multiplying for each pair of nodes the number of neighbours each one has.
    \item The number of triangles in which the nodes are. More precisely, the maximum number of triangles for a node and the minimum number of triangles.
    \item 	The clustering coefficient. Here we also have the maximal and minimum coefficient.
    \item The community detection by Leuven and Label Propagation (same Louvain and same Partition). This is simply a Boolean value that indicates whether two nodes are in the same community calculated by Louvain or by Label Propagation.
\end{enumerate}

The Scikit learn package in Python provides a function to display the importance of the different features in the model. Accuracy and memorization have been greatly improved, and the accuracy is still relatively high. The area under the curve is now very close to 1.

Finally, we have a model that makes predictions with an acceptable success rate. It does not perform as well as one would want to use it on a large scale. (In any case, it is a model that predicts interactions in a series of books, so the usefulness is quite limited). Nevertheless, it has the merit of having an acceptable performance for such a small data set. It also provides a method for finding good features and improving a link prediction model.

Here, if we wanted a simpler model but still quite efficient, we could have kept only the first model (see Figure \ref{fig:model}). characteristics.

\paragraph{Results.}
During training, we tested different characteristics, which seemed logical considering that our graph is a graph of the relationships among people. For example, two people who are not related but with a large number of neighbours. The characters in common seem to be more likely to interact in the future.

\begin{figure*}[h]
   \centering
   \includegraphics[width=\linewidth]{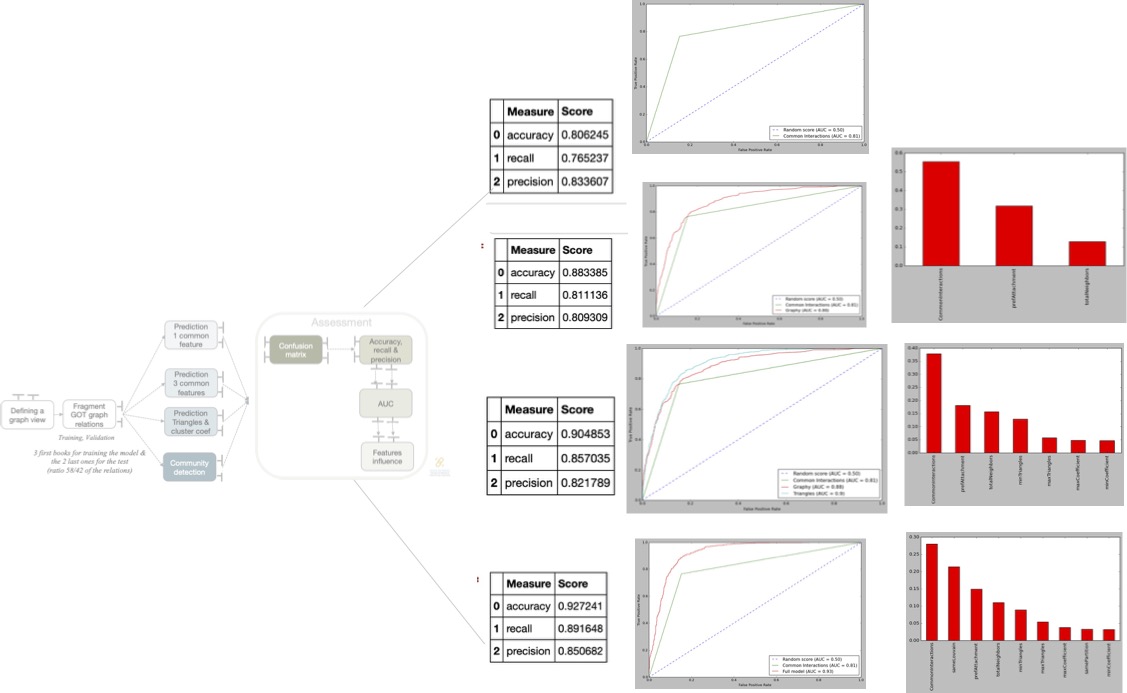}
    \caption{Link prediction models assessment}
   \label{fig:assessment}
 \end{figure*}

Figure \ref{fig:assessment} summarises the different assessment results with the different experimental settings tested for link prediction. These tasks correspond to the assessment part of our data science pipeline. The link prediction query could be defined in a general manner as a template since it was designed as an abstract pipeline. Then, different pipeline instances adopting different strategies for defining the graph view led to our experimental panel. 
\section{Conclusions and Future Work}\label{sec:conclusion}
This paper described and reported on an experimental comparative study to compare the imperative and declarative paradigms for programming data science pipelines on graphs. Imperative approaches rely on libraries and execution environments with no built-in options for managing graph views, resources allocation and graph persistence. In contrast, declarative approaches relying on underlying graph management systems profit from the manager's strategies for managing the graphs on disk and main memory. Our link prediction experiment showed that using the graph management system for creating views can be very elegant and sound; Then, given the cost of the algorithm, relying upon a parallel execution framework as Spark provides a more natural way of dealing with main memory allocation.

Based on these observations about graphs and other related work, our current work addresses the efficient execution of pipelines applied to the analysis of  graphs. 
We are deploying data science pipelines on target architectures such as the cloud and GPUs provide large-scale computing, memory and storage resources to further develop our experiments. There is room for querying and exploiting data through data science queries managed by the environment as first-class citizens for future work. 


\bibliographystyle{splncs04}
\bibliography{biblio}
\end{document}